\documentclass{article}
\usepackage{epsfig}
\usepackage{amsmath}

\date{\today} 
 
\usepackage{graphicx}

\newcommand{\insertplot}[5]{\begin{figure}
 \hfill\hbox to 0.05in{\vbox to #5in{\vfill
 \inputplot{#1}{#4}{#5}}\hfill}
 \hfill\vspace{-.1in}
 \caption{#2}\label{#3}
 \end{figure}}
 \newcommand{\inputplot}[3]{
 \special{ps: plotfile #1}
}
\newcounter{fig}   

\newcommand{\vphi}{\varphi}
\newcommand{\vepsilon}{\varepsilon}

\newcommand{\bx}{\bar{x}}
\newcommand{\beq}{\begin{equation}}
\newcommand{\eeq}{\end{equation}}
\newcommand{\beqs}{\begin{eqnarray}}
\newcommand{\eeqs}{\end{eqnarray}}

\numberwithin{equation}{section}

\usepackage{graphicx}

\begin{document}
\title{ Higher dimensional rotating black holes in 
Einstein-Maxwell theory \\with negative 
cosmological constant} 

\author{{\large Jutta Kunz}$^{1}$, 
{\large Francisco Navarro-L\'erida}$^{1}$%
  \ and {\large Eugen Radu}$^2$ 
\\
\\
$^{1}${\small Institut f\"ur Physik, Universit\"at Oldenburg, Postfach 2503
D-26111 Oldenburg, Germany}\\ 
$^{2}${\small Department of Mathematical Physics, National University of Ireland,
Maynooth, Ireland}}

\maketitle

\begin{abstract}
We present arguments for the existence of charged,
rotating black holes 
with equal-magnitude angular momenta in an odd number of dimensions  $D\geq 5$.
These solutions posses
a regular horizon of spherical topology and 
approach asymptotically the Anti-de Sitter spacetime background. 
We analyze their global charges, their gyromagnetic ratio
and their horizon properties.

\end{abstract}
 
\section{Introduction}
Recently a tremendous amount of interest has focused on Anti-de Sitter (AdS) spacetime.
This interest is mainly motivated by the proposed correspondence between physical effects
associated with gravitating fields propagating in AdS spacetime and those of a conformal
field theory (CFT) on the boundary of AdS spacetime \cite{Witten:1998qj,Maldacena:1997re}.

In this context, the  black holes  with  cosmological constant $\Lambda<0$
are of special interest since they would offer  the possibility  
of studying the nonperturbative structure of some CFTs.
Higher dimensional rotating black holes with AdS asymptotics have been studied by various authors,
starting with Hawking ${\it et. al.}$ \cite{Hawking:1998kw} 
who
generalized the $D=4$ Kerr-AdS solution to five dimensions 
with arbitrary angular momenta, and to
all dimensions with only a single nonzero angular momentum.
The generalization of these solutions to
the full set of independent angular momenta was given
in \cite{Gibbons:2004js}.

It is of interest to generalize these higher dimensional 
Kerr-AdS metrics further, by including matter fields.
Several exact solutions describing charged rotating black holes
have been found recently
in gauged supergravities in $D=5$  \cite{Chong:2005hr},
and $D=7$ \cite{Chong:2004dy} dimensions.
 Apart from Abelian fields with a Chern-Simons term, these configurations
usually contain scalar fields with a nontrivial scalar potential.

The main purpose of this paper
is to report progress on this problem by presenting numerical
evidence for the existence a set
of asymptotically AdS charged rotating black holes.
These configurations exist in an odd number of dimensions, $D \geq 5$.
They possess a regular horizon of spherical topology,
and their angular momenta are all of equal-magnitude,
thus factorizing the angular dependence.
The same approach was employed recently to construct
asymptotically flat charged rotating black holes in higher dimensions
\cite{Kunz:2006eh,Kunz:2005nm}. 

Also, instead
of specializing to a particular supergravity model, we shall consider
pure Einstein-Maxwell (EM) theory with negative cosmological
constant.
Although this theory is non-supersymmetric
in itself for $D>4$, it enters all gauged supergravities as the basic
building block. Therefore  one can expect the basic features of its solutions to be
generic.

The paper is structured as follows:  
in Section 2 we present the general   
framework and analyze the field equations.
The boundary conditions and the black hole properties
are discussed in Section 3.
We present the numerical results for $D=5$ and $D>5$ black holes
in Section 4,
and conclude with Section 5, 
where further applications are addressed.

\section{Action and Ansatz}
We consider the EM action with a 
negative cosmological constant $\Lambda$
\begin{eqnarray}
I=\frac{1}{16 \pi G_D}\int_\mathcal{M}~d^Dx \sqrt{-g} (R-2\Lambda-F_{\mu \nu}F^{\mu \nu})
-\frac{1}{8\pi G_D}\int_{\partial \mathcal{M}} d^{D-1}x\sqrt{-h}K, \label{action-grav}
\end{eqnarray}
where $D=2N+1$ ($N \ge 2$), $G_D$ is the $D$-dimensional Newton constant,
$R$ is the curvature scalar, and $F_{\mu  \nu}$ is the gauge field strength tensor
($F_{\mu \nu}=\partial_\mu A_\nu - \partial_\nu A_\mu$, with gauge potential
$A_\mu$). The last term in  (\ref{action-grav}) is the Hawking-Gibbons surface
term \cite{Gibbons:1976ue},
which is required in order to have a well-defined variational principle.
$K$ is the trace
of the extrinsic curvature for the boundary $\partial\mathcal{M}$ and
$h$ is the induced
metric of the boundary.
We denote $\Lambda=-(D-2)(D-1)/(2 \ell^2)$.

The field equations associated with the action (\ref{action-grav}) are the
Einstein equations
\begin{equation}
G_{\mu\nu}+\Lambda g_{\mu\nu} = 2 \left(F_{\mu\rho} {F_\nu}^\rho - \frac{1}{4}
g_{\mu \nu} F_{\rho  \sigma} F^{\rho \sigma}\right)
\ , \label{ee}
\end{equation}
and the gauge field equations
\begin{equation}
\nabla_\mu F^{\mu\nu}  = 0 \ .
\label{feqA}
\end{equation}
 
We consider stationary black hole space-times with $N$ azimuthal symmetries,
representing charged $U(1)$ generalizations of the corresponding set of 
vacuum solutions discussed in \cite{Gibbons:2004js}.
The symmetries imply the existence of $N+1$ commuting Killing vectors,
$\xi \equiv \partial_t$, 
and $\eta_{(k)} \equiv \partial_{\vphi_k}$, for $k=1, \dots , N$. While the
general EM-AdS black holes then possess $N$ independent
angular momenta, we here restrict to black holes 
with equal-magnitude angular momenta 
and with spherical horizon topology\footnote{ 
Asymptotically AdS rotating charged topological black hole
solutions with zero scalar curvature  of the
event horizon are known in closed
form \cite{Awad:2002cz},
but they were found for a different metric ansatz, and they possess rather
different properties.}.

We employ a parametrization for the metric
corresponding to a generalization of the Ansatz
used previously for asymptotically flat solutions \cite{Kunz:2006eh}.
It has the general form
\begin{eqnarray}
 ds^2 = -b(r)dt^2 +  \frac{ dr^2}{u(r)} +
g(r)\sum_{i=1}^{N-1}
  \left(\prod_{j=0}^{i-1} \cos^2\theta_j \right) d\theta_i^2
 +p(r) \sum_{k=1}^N \left( \prod_{l=0}^{k-1} \cos^2 \theta_l
  \right) \sin^2\theta_k \left( \vepsilon_k d\vphi_k - \frac{w(r)}{r}
  dt\right)^2
\nonumber
 \\
 +(g(r)-p(r)) \left\{ \sum_{k=1}^N \left( \prod_{l=0}^{k-1} \cos^2
  \theta_l \right) \sin^2\theta_k  d\vphi_k^2 \right.
  -\left. \left[\sum_{k=1}^N \left( \prod_{l=0}^{k-1} \cos^2
  \theta_l \right) \sin^2\theta_k  \vepsilon_k d\vphi_k\right]^2 \right\} \ ,
  \label{metric}
\end{eqnarray}
where $\theta_0 \equiv 0$, $\theta_i \in [0,\pi/2]$
for $i=1,\dots , N-1$,
$\theta_N \equiv \pi/2$, $\vphi_k \in [0,2\pi]$ for $k=1,\dots , N$,
and $\vepsilon_k = \pm 1$ denotes the sense of rotation
in the $k$-th orthogonal plane of rotation. 
For such solutions, the isometry group is enhanced from $R \times U(1)^{N+1}$
to $R \times U(N+1)$, where $R$ denotes the time translation.
This symmetry enhancement allows us to deal only with ordinary differential
equations (ODE's).

The vacuum black holes discussed in \cite{Gibbons:2004js}
are recovered for vanishing gauge field and
\begin{eqnarray}
\label{vacuum}
 u(r)=1
+\frac{r^2}{\ell^2}
-\frac{2{\hat M}\Xi}{r^{D-3}}
+\frac{2{\hat M}{\hat a}^2}{r^{D-1}},~
p(r)=r^2\left(1+\frac{2{\hat M}{\hat a}^2}{r^{D-1}}\right),~
w(r)=\frac{2{\hat M}{\hat a}}{r^{D-4}p(r)},~
g(r)=r^2,~ b(r)=\frac{r^2u(r)}{p(r)},
\end{eqnarray}
where ${\hat M}$ and ${\hat a}$ are two constants related to the solutions' mass and 
angular momentum, and $\Xi=1-{\hat a}^2/\ell^2$ (see 
\cite{Kunduri:2006qa} for a discussion of the basic
features of these solutions).
For the numerical calculations
a convenient parametrization is given by
\begin{eqnarray}
\label{proposal} 
u(r)=\frac{f(r)}{m(r)}\left(\frac{r^2}{\ell^2}+1\right),
~~b(r)=\left(\frac{r^2}{\ell^2}+1\right)f(r),~~g(r)=\frac{m(r)}{f(r)}r^2,
~~p(r)=\frac{n(r)}{f(r)}r^2.
\end{eqnarray}

The Ansatz for the U(1) potential, consistent with the symmetries
of the  line element (\ref{metric}), is given by \cite{Kunz:2006eh}
\begin{equation}
A_\mu dx^\mu =  a_0 dt + a_\vphi \sum_{k=1}^N \left(\prod_{l=0}^{k-1}
  \cos^2\theta_l\right) \sin^2\theta_k \vepsilon_k d\vphi_k \ .
\label{gaugepotential}
\end{equation}
Thus, independent of the odd dimension $D\ge 5$,
this parametrization involves only four functions $f, m, n, \omega$
for the metric and two functions $a_0, a_\vphi$
for the gauge field, which all depend only on the radial coordinate $r$.

When (\ref{metric}) and (\ref{gaugepotential}) are substituted into (\ref{ee})
and (\ref{feqA}), a coupled system of ODE's is obtained (one first-order equation (for
$n$) plus five second-order equations (for $f, m, \omega, a_0,
A_\vphi$)). However, taking advantage of the existence of a first integral of
that system
\begin{equation}
\frac{r^{D-2} m^{(D-5)/2}}{f^{(D-3)/2}} \sqrt{\frac{m n}{f}} \left(\frac{d
      a_0}{dr} + \frac{\omega}{r} \frac{d a_\vphi}{dr} \right) = (D-3) q \ , \
      \ \
       q={\rm constant} \label{first_integral} \ ,
\end{equation}
we may eliminate $a_0$ from the equations, leaving a 
system of one first order equation (for $n$) and
four second order equations \cite{Kunz:2006eh}.

\section{Black Hole Properties}
\subsection{Asymptotic expansion and boundary conditions}

In order to generate black hole solutions which are asymptotically  AdS 
and possess a regular event horizon of spherical topology, we have to impose
 appropriate boundary conditions. 

A straightforward computation
 gives the following asymptotic 
expansion for the metric and matter fields, involving five 
parameters ${\tilde \alpha},~{\tilde \beta},~ {\hat J},~q$ and $\hat \mu$,
\begin{eqnarray}
&&f= 1+\frac{{\tilde \alpha}}{r^{D-1}} + O\left(\frac{1}{r^{D+1}}\right) \ ,
~~
m=1+\frac{{\tilde \beta}}{r^{D-1}} + O\left(\frac{1}{r^{D+1}}\right) \ ,
~~
n= 1+\frac{(D-2)({\tilde \alpha}-{\tilde \beta})}{r^{D-1}} + 
O\left(\frac{1}{r^{D+1}}\right) \ ,
\nonumber
\\
&&\omega= \frac{{\hat J}}{r^{D-2}} + O\left(\frac{1}{r^{2D-5}}\right) \ ,
~~
a_0= -\frac{q}{r^{D-3}} + O\left(\frac{1}{r^{2D-4}}\right) \ ,
~~
a_\vphi= \frac{{\hat \mu}}{r^{D-3}} + O\left(\frac{1}{r^{2D-4}}\right) \ . 
\label{exp_inf}
\end{eqnarray}
Therefore,  in the numerical procedure, we impose  the following boundary conditions
at infinity 
\begin{equation}
f|_{r=\infty}=m|_{r=\infty}=n|_{r=\infty}=1 \ , \ \omega|_{r=\infty}=0 \ , 
\ a_0|_{r=\infty}=a_\vphi|_{r=\infty}=0 \ .
\label{bc1} \end{equation}

The horizon of these black hole solutions is a
 squashed $S^{D-2}$ sphere.
It resides at the constant value of the radial coordinate $r=r_{\rm
  H}$, and is characterized by $f(r_{\rm H})=0$.
Expanding the metric and matter functions at the horizon yields
\begin{eqnarray}
\nonumber
&&f=f_2 \bx^2\left(1-\frac{2 r^2_{\rm H} -\ell^2}{r^2_{\rm H}+\ell^2} \bx \right) +
O(\bx^4) \ , ~~
m=m_2 \bx^2\left(1-\frac{4 r^2_{\rm H} +\ell^2}{r^2_{\rm H}+\ell^2} \bx \right) +
O(\bx^4) \ , \label{exp_hor}
\\
&&n=n_2 \bx^2\left(1-\frac{4 r^2_{\rm H} +\ell^2}{r^2_{\rm H}+\ell^2} \bx \right) +
O(\bx^4) \ ,~~
\omega= \Omega  r_{\rm H} (1+\bx) + O(\bx^2) \ , 
\\
\nonumber
&&a_0= -(\Phi_{\rm H}+\Omega a_{\vphi_0}) +  O(\bx^2) \ , ~~
a_\vphi=  a_{\vphi_0} +  O(\bx^2) \ , 
\end{eqnarray}
(with $f_2,m_2,n_2$ positive constants).
Here $\Omega$ is the (constant) horizon angular velocity defined  in terms of
the Killing vector
\begin{equation}
\label{chi}
\chi = \partial_t + \Omega \sum_{k=1}^N \vepsilon_k \partial_{\vphi_k} \ ,
\end{equation}
which is null at the horizon. $\Phi_{\rm H}$ denotes the (constant) horizon
 electrostatic potential, and the compactified radial coordinate $\bx$ is given
by $\bx = 1-r_{\rm H}/r$.

At the horizon, the solutions satisfy the boundary conditions 
\begin{eqnarray}
&&f|_{r=r_{\rm H}}=m|_{r=r_{\rm H}}=n|_{r=r_{\rm H}}=0 \ ,
\ \omega|_{r=r_{\rm H}}=r_{\rm H} \Omega \ , \ \nonumber \\ 
&&\Phi_{\rm H} = - \left. (a_0+\Omega a_\vphi)\right|_{r=r_{\rm H}} \ , \ 
\left. \frac{d a_\vphi}{d r}\right|_{r=r_{\rm H}}=0 \ . 
\label{bc2} 
\end{eqnarray}

\subsection{Global charges}


In an asymptotically flat spacetime,
the total mass $M$ and the angular momenta $J_{(k)}$ of the black holes
are obtained from the Komar expressions
associated with the respective Killing vector fields.
For AdS asymptotics,
the calculation of the total mass is more involved,
mainly because the analogous Komar integral for the relevant
time-like Killing vector diverges, which then requires
a regularization. Various definitions of the conserved charges
for an AdS background have been presented in the
literature  \cite{Hollands:2005wt}.

Employing first the Ashtekar-Magnon-Das conformal mass definition \cite{Ash},
we obtain for the mass of rotating EM black holes the expression
\begin{eqnarray}
\label{mass}
M=-\frac{A(S^{D-2})}{16\pi G_D}~ \frac{{\tilde \beta} + (D-2) {\tilde
    \alpha}}{\ell^2} \ , 
\end{eqnarray}
where $A(S^{D-2})$ denotes the area of the unit $(D-2)$-sphere,
and ${\tilde \alpha},~{\tilde \beta}$  the constants in the asymptotic
expansion (\ref{exp_inf}).

A different technique, proposed by Balasubramanian
and Kraus \cite{Balasubramanian:1999re},
was inspired by the AdS/CFT correspondence and consists 
in adding suitable counterterms $I_{ct}$
to the action of the theory in order to ensure the finiteness of the boundary
stress tensor 
${\rm T}_{ab}= \frac{2}{\sqrt{-h}} \frac{\delta I}{ \delta h^{ab}}$
derived by the quasilocal energy definition
\cite{Brown:1993br}. 
The expression for the mass we obtain
in this approach\footnote{
We have computed the mass and angular momenta for solutions in 
$D=5,~7$ and 9 dimensions by using the
 expressions of the counterterms and 
boundary stress-tensor given $e.g.$ in  Ref.~\cite{Das:2000cu}. 
If there are matter fields on $\cal{M}$, additional counterterms
may be needed to regularize the action and global charges.
However, we have found that for a purely Abelian matter content, 
the usual counterterm prescription removes all divergences.} 
contains besides the expression (\ref{mass}) for the mass
an additional term $E_c^{(D)}$,
which depends only on $\ell$ and $D$
($e.g.$ $E_c^{(5)}=3 A(S^3)\ell^2/64\pi G_5,~ E_c^{(7)}=- 5 A(S^5)\ell^4/128\pi G_7$).
$E_c^{(D)}$ corresponds to the mass of the pure global AdS$_{2N+1}$ and is usually
interpreted as the energy dual to the Casimir energy of the dual CFT defined on
the $(D-1)$-dimensional boundary metric \cite{Balasubramanian:1999re}.

The angular momenta $J_{(k)}$ may be computed from the standard Komar integral,
 since the
divergent term vanishes. The Komar integral reads 
\begin{equation}
J_{(k)} = \frac{1}{16 \pi G_D}  \int_{S_{\infty}^{D-2}} \beta_{(k)} \ ,
\end{equation}
with $\beta_{ (k) \mu_1 \dots \mu_{D-2}} \equiv \epsilon_{\mu_1 \dots \mu_{D-2}
  \rho \sigma} \nabla^\rho \eta_{(k)}^\sigma$,
and for equal-magnitude angular momenta $J_{(k)}=\varepsilon_k J$, 
$k=1, \dots , N$. With (\ref{exp_inf}) we then obtain 
  for the angular momentum the expression
\begin{equation}
J=\frac{A(S^{D-2})}{8\pi G_D} {\hat J} \ ,
 \label{ang_mom}
\end{equation}
which agrees with the value found 
within the counterterm approach\footnote{ There is ostensibly a 
mismatch between the total number of parameters 
in the asymptotic expansion (\ref{exp_inf})
and the number of conserved charges of the solution, which we could
not clarify.
For $Q=0$, one finds ${\hat J}= \sqrt{(2\tilde\alpha-\tilde\beta)
((D-1)\tilde\beta-(D-2)\tilde\alpha)}/\ell$, while
$\tilde\beta=\tilde\alpha (D-2)/(D-1)$ in the static limit.
%
However, 
for charged rotating solutions,
the numerical results do not indicate the existence of any simple correlation
 between the constants 
 ${\tilde \alpha},~{\tilde \beta},~{\hat J}$ and $q$ in (\ref{exp_inf}).

}.

The electric charge is related to the first integral (\ref{first_integral}) by
\begin{equation}
Q=\frac{(D-3) A(S^{D-2})}{4\pi G_D} q \ , \label{elec_charge}
\end{equation}
and the magnetic moment $\mu_{\rm mag}$ is given by
\begin{equation}
\mu_{\rm mag} = \frac{(D-3) A(S^{D-2})}{4\pi G_D} {\hat \mu} \ . \label{mag_moment}
\end{equation}

A further quantity of physical interest is the gyromagnetic ratio 
$g$ of these charged
rotating AdS black holes, defined as
\begin{equation}
g=\frac{2 M \mu_{\rm mag}}{Q J} \ . 
\label{g_factor}
\end{equation}

\subsection{Horizon properties}

Employing the expansion at the horizon (\ref{exp_hor}),
we obtain for the surface gravity $\kappa$, 
 defined by
\begin{equation}
\kappa^2 =\left. -\frac{1}{2}  (\nabla_\mu \chi_\nu) (\nabla^\mu \chi^\nu)
\right|_{r_{H}} \ , \label{sg1}
\end{equation}
the expression
\begin{equation}
\kappa=\left(1+\frac{r^2_{\rm H}}{\ell^2}\right) \frac{f_2}{r_{\rm H} \sqrt{m_2}}
\ , \label{sg2}
\end{equation}
and for the horizon area 
\begin{equation}
A_{\rm H} = r^{D-2}_{\rm H} A(S^{D-2}) \sqrt{\frac{m^{D-3}_2
    n_2}{f^{D-2}_2}} \ . \label{hor_area}
\end{equation}
To have a measure of the deformation of the horizon, we introduce a
deformation parameter defined as the ratio of the equatorial circumference
$L_e$ and
the polar one $L_p$, which for these solutions we are considering takes the
form
\begin{equation}
L_e/L_p = \sqrt{\frac{n_2}{m_2}} \ . \label{L_ratio}
\end{equation}

The horizon mass $M_{\rm H}$ and the horizon angular momentum
$J_{\rm H}$ can be computed by evaluating the corresponding Komar integrals at
the horizon \cite{Kunz:2006eh}, leading to 
\begin{eqnarray}
&&M_{\rm H} = \frac{1}{8\pi G_D} \frac{D-2}{D-3} A(S^{D-2}) r^{D-3}_{\rm H}
\sqrt{\frac{m^{D-4}_2 n_2}{f^{D-4}_2}}  \left[1 + \frac{ r^2_{\rm H}}{\ell^2} -
  \frac{n_2}{2 f^2_2} r_{\rm H} \Omega \left(\left.\frac{d^2\omega}{d \bx^2}
    \right|_{\bx=0} -2  r_{\rm H} \Omega \right) \right]  \ , \nonumber \\
&&J_{\rm H} = \frac{1}{8\pi G_D (D-1)} A(S^{D-2}) r^{D-2}_{\rm
  H} \sqrt{\frac{m^{D-4}_2 n^3_2}{f^{D}_2}} \left( 2  r_{\rm H} \Omega -
  \left.\frac{d^2\omega}{d \bx^2}    \right|_{\bx=0} \right)  \ . \label{hor_MJ}\end{eqnarray}
These horizon quantities are related by the horizon mass formula
\begin{equation}
\frac{D-3}{D-2} M_{\rm  H} = \frac{\kappa A_{\rm H}}{8 \pi G_D} + N \Omega
J_{\rm H} \ . \label{hor_mass_formula}
\end{equation}
 To obtain the global counterpart of (\ref{hor_mass_formula}) remains
 a challenge.
It would be interesting to extend the  approach used in \cite{Barnich:2004uw},
to derive a Smarr-type formula for Kerr-AdS solutions,
to the case discussed in this paper.

\subsection{Ergosurface and closed causal curves}

These rotating black holes possess an ergosurface,
inside of which
observers cannot remain stationary, but will 
move in the direction of the rotation. 
The ergosurface is located at a constant value of the
radial coordinate $r=r_e$, with $g_{tt}(r_e)=0$, $i.e.$
\begin{eqnarray} 
\label{er}  
\frac{n(r_e)}{f(r_e)}w^2(r_e)- f(r_e) \left(1+\frac{r^2_e}{\ell^2}\right)=0~,
\end{eqnarray} 
and does not intersect the horizon. 

Also, as a result of the metric ansatz (\ref{metric}),
$F(x^\mu)=t$ is a global time coordinate ($g^{\mu \nu}  F_{,\mu}F_{,\nu} 
=-1/(f(r)(r^2/\ell^2+1))<0$),
and no closed causal curves occur in the region outside the event horizon.

\section{Results}

The numerical methods employed here are analogous to those used to construct
asymptotically flat charged rotating black holes \cite{Kunz:2006eh}. 
Working
with the compactified coordinate $\bar x $, we choose units such that $G_D=1$,
and apply a collocation method for boundary-value ordinary
differential equations, equipped with an adaptive mesh selection procedure
\cite{COLSYS}.
Typical mesh sizes include $10^3-10^4$ points.
The solutions have a relative accuracy of $10^{-8}$. 
 
In the following we focus on the properties of 
 $D=5$ black hole solutions. These results can be easily
extended to odd dimensions $D>5$,
since the main features of the solutions are common for all odd dimensions. 
We note, that the computation of the mass involves
$1/r^{D-1}$ terms in the asymptotic expansion, 
decreasing the numerical accuracy of the mass as $D$ increases.

\subsection{Domain of existence}

Let us first address the domain of existence of these black hole
solutions. In order to do so, we note that the solutions have a scale symmetry,
which is broken once the value of the cosmological constant is fixed: 
 $\tilde \ell =\lambda \ell$,
 $\tilde M= \lambda^{D-3} M$,
 $\tilde J= \lambda^{D-2} J$,
 $\tilde Q = \lambda^{D-3} Q$,
 $\tilde r_{\rm H}=\lambda r_{\rm H}$,
 $\tilde \Omega = \Omega/\lambda$,
 $\tilde \kappa = \kappa/\lambda$, etc.
To classify solutions related by this symmetry, we introduce the scale
invariant ratio
\begin{equation}
\ell_Q=\frac{\ell}{|Q|^{1/(D-3)}} \ . 
\label{L_Q_def}
\end{equation}

\begin{figure}[t!]
\parbox{\textwidth}
{\centerline{
\mbox{
\epsfysize=10.0cm
\includegraphics[width=87mm,angle=0,keepaspectratio]{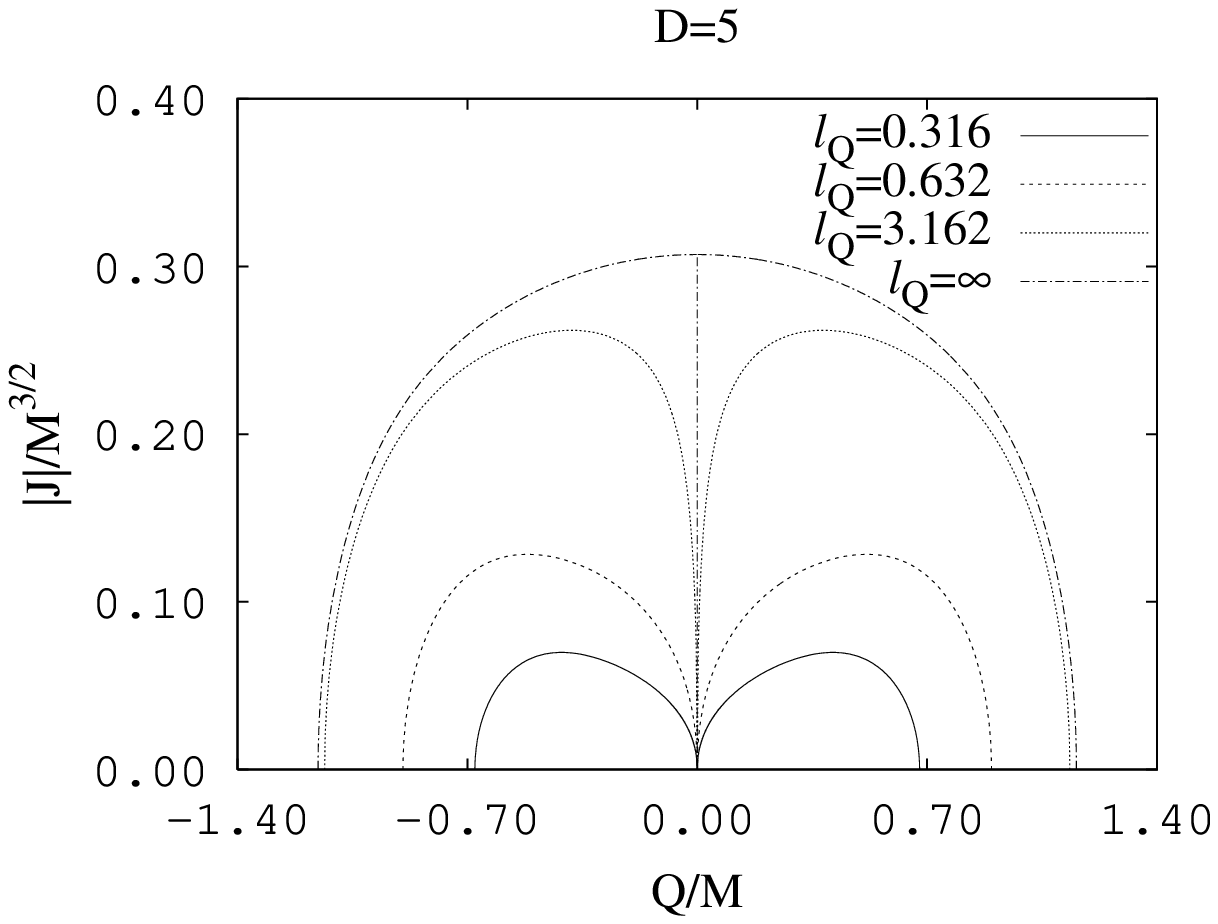}
\includegraphics[width=87mm,angle=0,keepaspectratio]{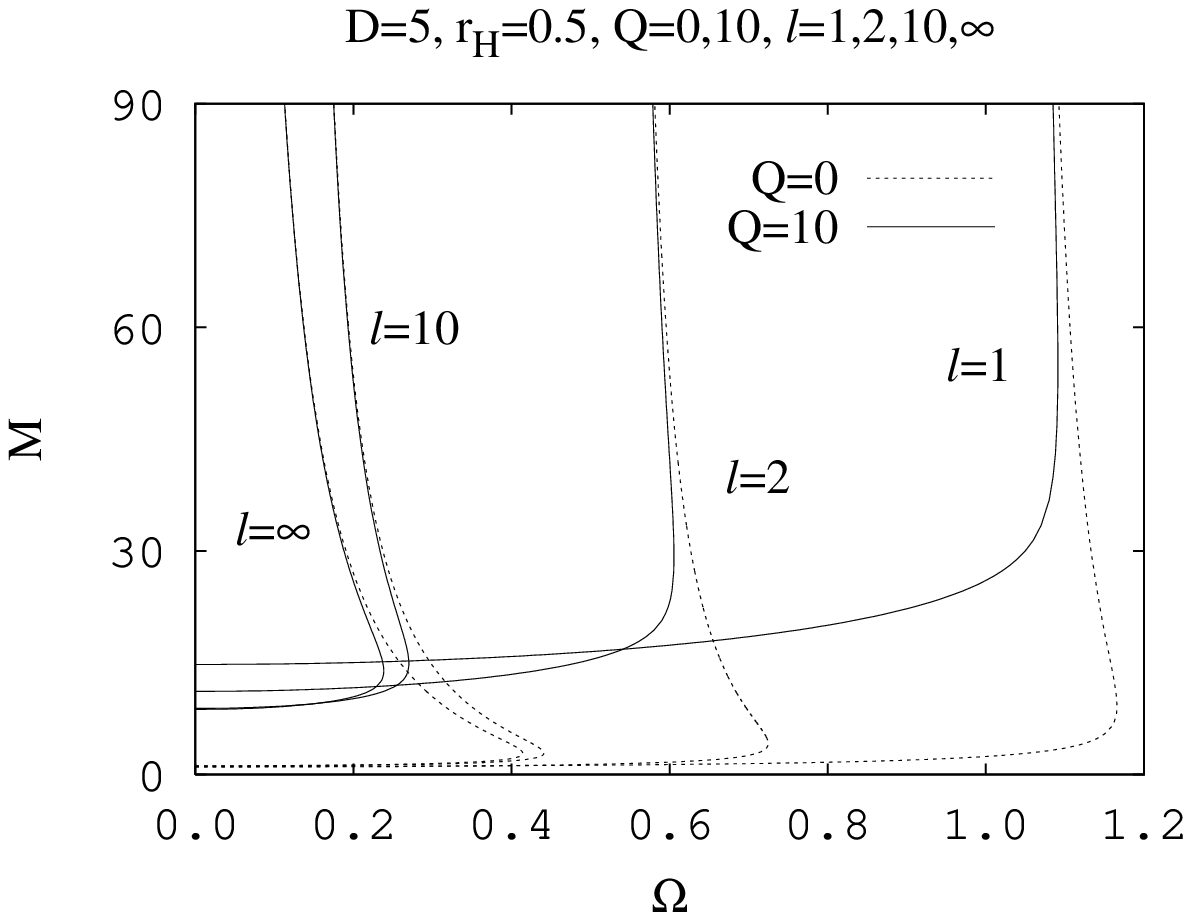}
}}}
\caption{{\small Left: Scaled angular momentum $J/M^{(D-2)/(D-3)}$ 
$vs.$~scaled charge $Q/M$ for extremal black holes
with equal-magnitude angular momenta
in $D=5$ dimensions. 
Right: Mass $M$ $vs.$~horizon angular velocity $\Omega$
for equal-magnitude angular momenta black hole solutions
in $D=5$ dimensions
for fixed isotropic horizon radius $r_{\rm H}=0.5$, and
fixed charge $Q=10$, 
for several values of $\ell$; the uncharged Kerr-AdS counterparts are also
shown.}}
\end{figure}

In Fig.~1 (left) we show the domain of existence of $D=5$ solutions for several
values of $\ell_Q$. The domain of existence
is bounded by extremal solutions, which are
characterized by a vanishing surface gravity $\kappa$. For finite $\ell_Q$ the
extremal curve that delimits the corresponding region of existence reaches the
$|J|/M^{(D-2)/(D-3)}=0$ axis at $|Q|/M=0$ and at a finite value of $|Q|/M$,
affiliated to the corresponding extremal higher dimensional
Reissner-Nordstr\"om (RN)-AdS solutions.\footnote{
The analytical RN-AdS values $Q/M=0.677, 0.897, 1.135, 2/\sqrt{3}$
for $\ell_Q=0.316,0.632, 3.162, \infty$, respectively,
in Fig.~1 (left) agree accurately with the numerical values.}
As $\ell_Q$ increases, the origin is approached more and more
steeply and, in the
limit $\ell_Q=\infty$, the extremal curve consists of two pieces: the pure EM
part (without cosmological constant) \cite{Kunz:2006eh}, plus a vertical line
corresponding to uncharged Kerr-AdS solutions.
Clearly, the domain of
existence enlarges with increasing $\ell_Q$,
and is symmetric with respect to $Q \to -Q$.

\boldmath
\subsection{Global charges and $g$-factor}
\unboldmath

We now turn to non-extremal black holes, and analyze their
asymptotic properties.
We first consider the global charges of $D=5$ black hole solutions,
obtained by keeping the charge $Q$ 
and the isotropic horizon radius $r_{\rm H}$ fixed,
while varying the horizon angular velocity $\Omega$.
The mass $M$ of these solutions as given by (\ref{mass}) is exhibited in Fig.~1 (right),
together with the mass of their higher dimensional
Kerr-AdS counterparts.

For each set of solutions we observe two branches,
extending up to a maximal value of $\Omega$,
where they merge and end.
The lower branch emerges from the static solution in the limit $\Omega=0$.
On the upper branch the mass diverges in the limit 
$|\Omega| \rightarrow 1/\ell$. 
Thus a static horizon ($\Omega=0$) cannot be reached along the upper
branch for finite $\ell$.
The maximal value of $\Omega$
depends on the horizon radius $r_{\rm H}$, the charge $Q$, the length $\ell$,
and the dimension $D$.

The angular momentum $J$ of the solutions exhibits a very similar
dependence on $\Omega$ and $\ell$ as the mass.
We note that, for a fixed value of the charge, its influence 
on the solutions, and thus their deviation from the corresponding
higher dimensional Kerr-AdS solutions, 
decreases with increasing dimension $D$,
as expected from the scaling properties of the solutions.

\begin{figure}[h!]
\parbox{\textwidth}
{\centerline{
\mbox{
\epsfysize=10.0cm
\includegraphics[width=87mm,angle=0,keepaspectratio]{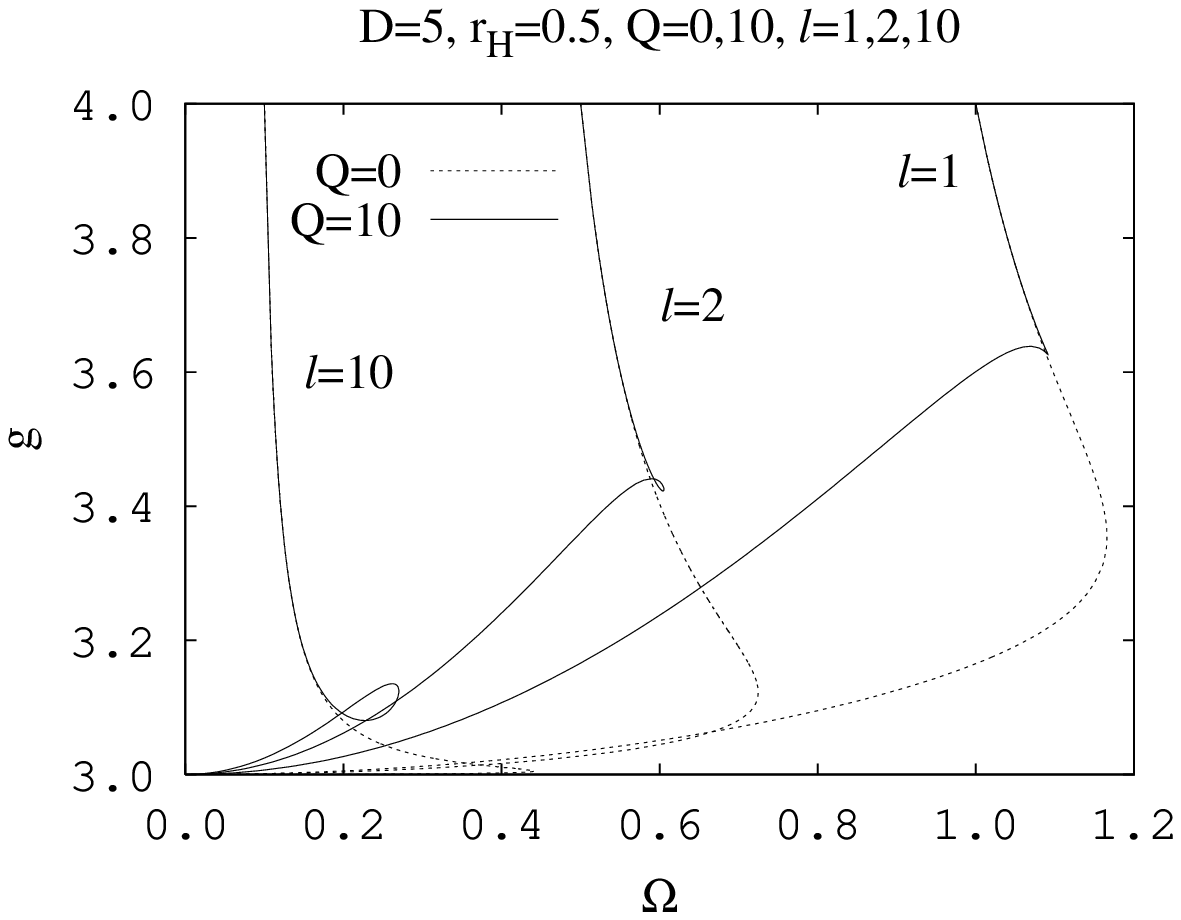}
\includegraphics[width=87mm,angle=0,keepaspectratio]{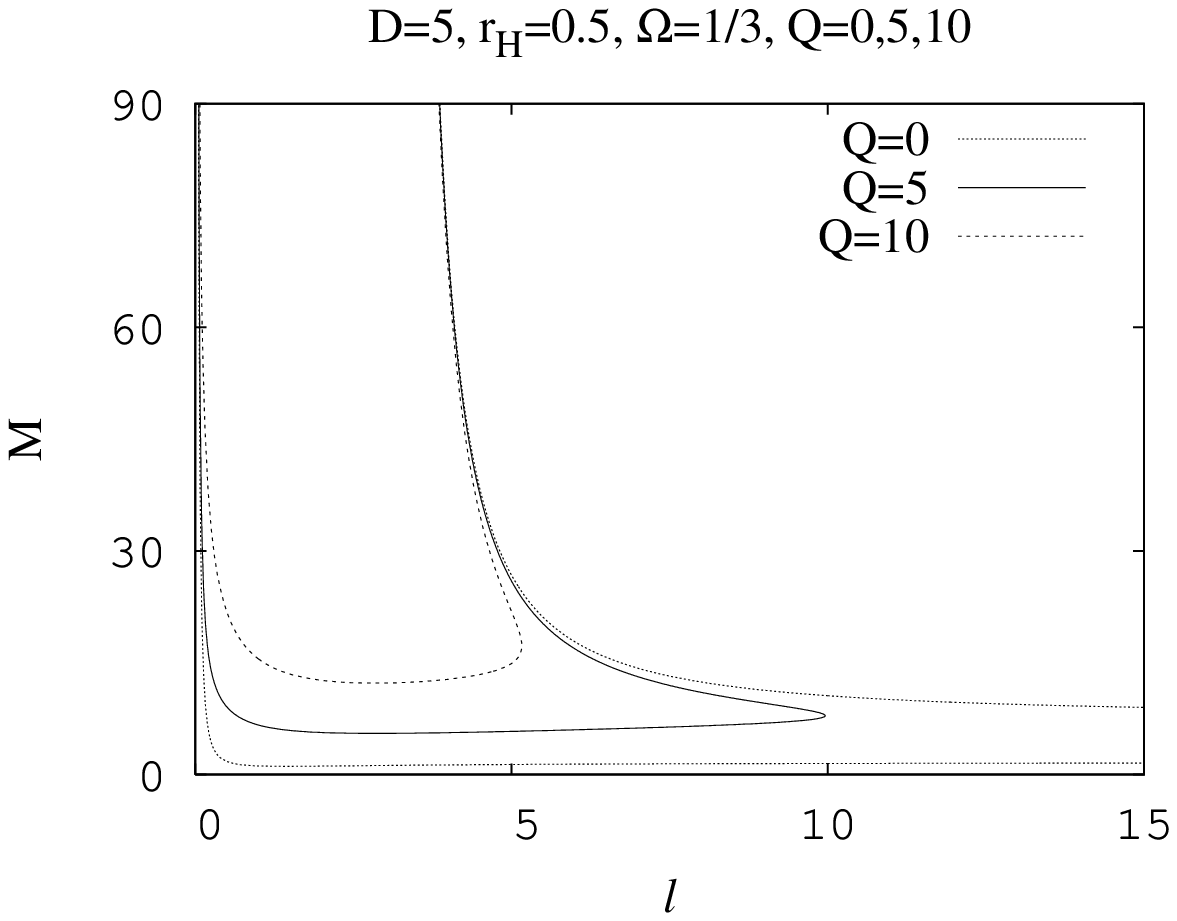}
}}}
\caption{{\small 
Left: Gyromagnetic ratio $g$ $vs.$~horizon angular velocity $\Omega$
for the same set of solutions as in Fig.~1 (right).
Right: Mass $M$ $vs.$~$\ell$
for equal-magnitude angular momenta black hole solutions
in $D=5$ dimensions
for fixed horizon radius $r_{\rm H}=0.5$,
angular velocity $\Omega=1/3$, and varying $\ell$ for several
values of the charge $Q=0, 5, 10$.}}
\end{figure}

The gyromagnetic ratio $g$ of these solutions is exhibited in Fig.~2 (left). 
It reveals clearly,
that $1/\ell$ is the limiting value of $\Omega$ on the upper branch. 
We note that the $g$-factor ranges between $D-2$ and $D-1$ 
for finite values of $\ell$, 
the value $D-1$ being reached along the upper branch,
when $|\Omega| \to 1/\ell$. 

In the limit $Q \to 0$, the gyromagnetic ratio 
can be obtained analytically by linearly perturbing the higher dimensional
Kerr-AdS solutions. 
We obtain for weakly charged solutions (in the limit $Q \to 0$),
possessing equal-magnitude angular momenta,
\begin{equation}
g= (D-2)+\frac{{\hat a}^2}{\ell^2} \ , \label{g_fact_perturb}
\end{equation}
with
\begin{equation}
M=\frac{A(S^{D-2})}{8 \pi G_D} \left[ (D-2) +\frac{{\hat a}^2}{\ell^2}\right]{\hat M} \ , \ \ \
J=\frac{A(S^{D-2})}{4 \pi G_D} {\hat M}{\hat a} \ .
\end{equation}
This shows clearly, that $g$ ranges between $D-2$ and
$D-1$ also for weakly charged solutions, since ${\hat a}^2/{\ell^2} \leq 1$.
We note, that for weakly charged solutions 
with a single non-vanishing angular momentum
the $g$-factor ranges between $D-2$ and 2 \cite{Aliev}.

In Fig.~2 (right) we exhibit the dependence of the mass on $\ell$,
for fixed $\Omega$ and $r_{\rm H}$, and several values of $Q$.
For fixed (but not too large) $\Omega$, 
there is just a single solution when $\ell<1/|\Omega|$. 
When $\ell>1/|\Omega|$, 
depending on the value of the charge $Q$,
there may be a maximal value of $\ell$ 
beyond which no solutions exist. 
As $Q$ increases, this maximal value of $\ell$ decreases. 
We note that this pattern is in agreement with Fig.~1 (right).

\subsection{Horizon properties}

Let us now address the horizon properties
of the solutions, beginning with the surface gravity $\kappa$
and the horizon area $A_{\rm H}$.
For most of the sets of solutions 
with fixed $r_{\rm H}$, $Q$ and $\ell$ and varying $\Omega$,
$\kappa$ decreases monotonically,
reaching a vanishing value at $|\Omega|=1/\ell $,
on the upper branch.
At the same time, the horizon area $A_{\rm H}$ increases monotonically,
diverging in the limit.

Complementary information on these solutions is 
obtained from Fig.~3 (left),
where the dependence of the area on the isotropic horizon radius $r_{\rm H}$ 
is exhibited, for fixed $\Omega$, $Q$, and $\ell$. 
Again we see, that $|\Omega|=1/\ell$ marks the borderline,
concerning the existence of one or two solutions. 
For $|\Omega| \leq 1/\ell$, there is just a single solution for each value
of  $r_{\rm H}$, and $r_{\rm H}$ can be increased without bound. 
In addition, as  $r_{\rm H}$ is increased, 
the difference between charged and uncharged solutions decreases. 
Beyond $|\Omega| = 1/\ell$ and below a maximal value of $\Omega$ 
(given by the corresponding maximal value of $\Omega$ for the extremal
solution), two solutions exist.

\begin{figure}[h!]
\parbox{\textwidth}
{\centerline{
\mbox{
\epsfysize=10.0cm
\includegraphics[width=87mm,angle=0,keepaspectratio]{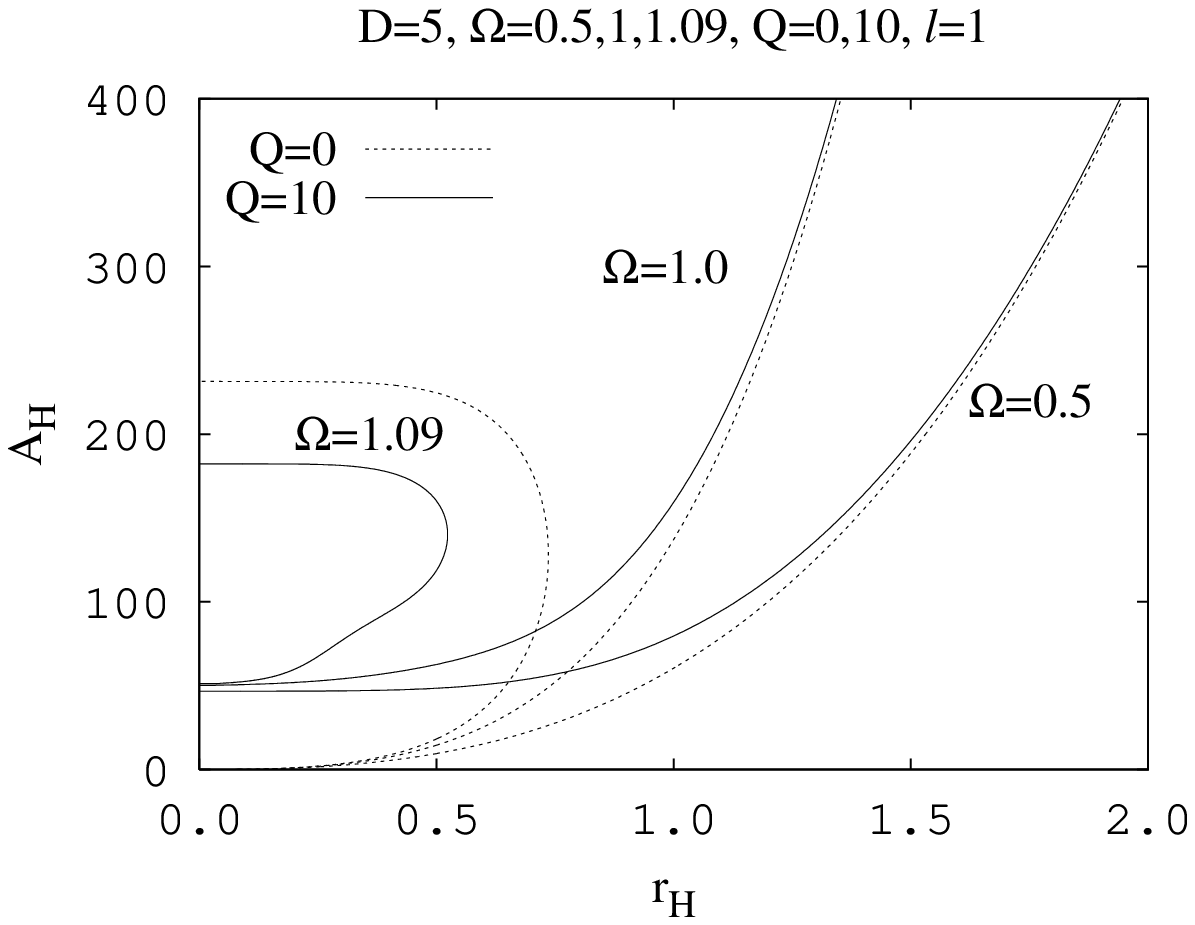}
\includegraphics[width=87mm,angle=0,keepaspectratio]{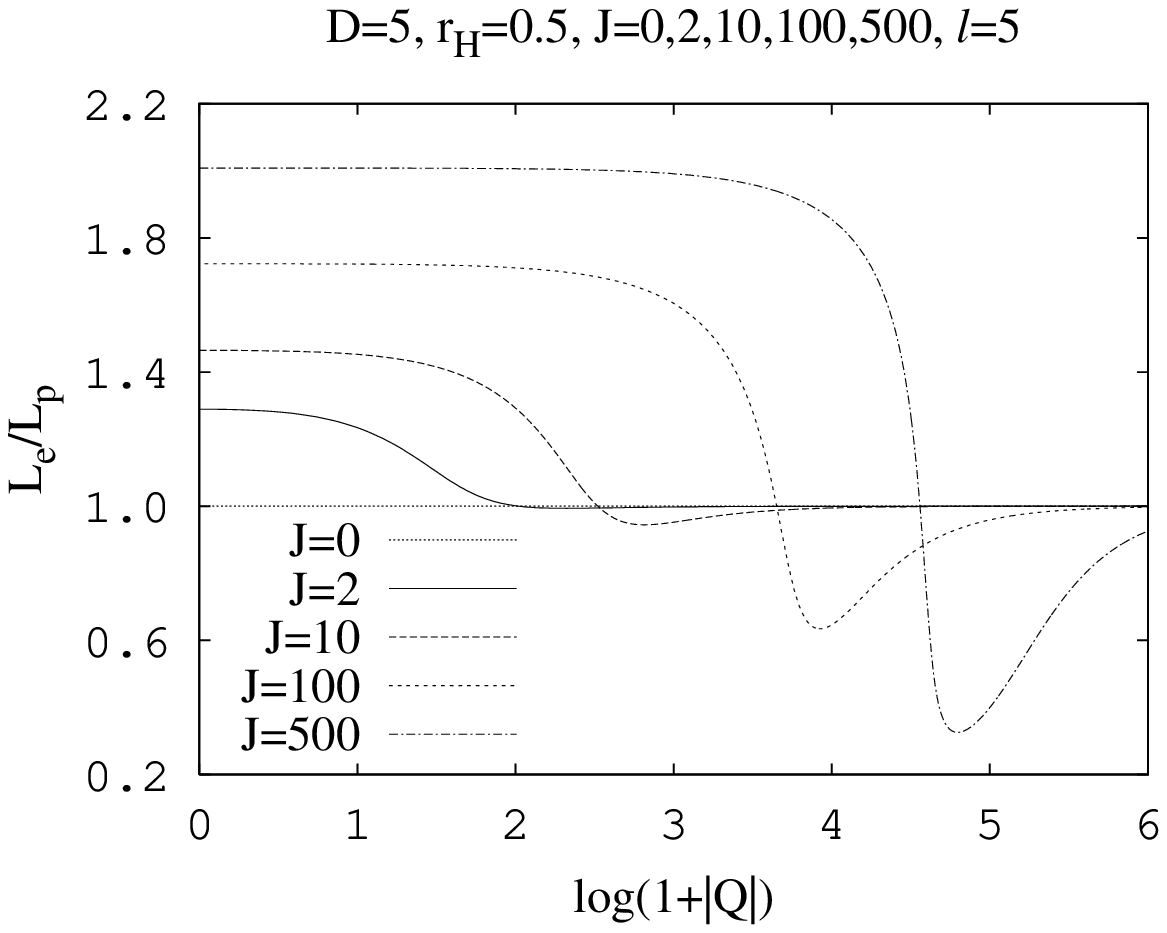}
}}}
\caption{{\small 
Left: Horizon area $A_{\rm H}$ $vs.$~isotropic
horizon radius $r_{\rm H}$ 
for equal-magnitude angular momenta black hole solutions in $D=5$ dimensions
for fixed charge $Q=10$, $\ell=1$, and
several values of the horizon angular velocity $\Omega=0.5, 1.0, 1.09$; their
uncharged counterparts are also shown.
Right: Ratio of horizon circumferences $L_{\rm e}/L_{\rm p}$
$vs.$~charge $Q$ 
for equal-magnitude angular momenta black hole solutions in $D=5$ dimensions
for fixed $r_{\rm H}=0.5$, $\ell=5$, and
several values of the angular momentum $J=0$, 2, 10, 100, 500.
}}
\end{figure}

To have a measure for the deformation of the horizon, we consider
the ratio of equatorial and polar circumferences
$L_{\rm e}/L_{\rm p}$ of the horizon.
(The equatorial circumference $L_{\rm e}$ 
is obtained with $\theta=0$ and constant $\varphi_1$,
and the polar circumference $L_{\rm p}$ 
with constant $\varphi_1$ and $\varphi_2$.)
Whereas Kerr-AdS solutions with equal angular momenta 
always have oblate deformation with
ratio $L_{\rm e}/L_{\rm p} \ge 1$, we observe in Fig.~3 (right),
that the presence of charge allows for prolate deformation
of the horizon in certain regions of parameter space.

\begin{figure}[h!]
\parbox{\textwidth}
{\centerline{
\mbox{
\epsfysize=10.0cm
\includegraphics[width=87mm,angle=0,keepaspectratio]{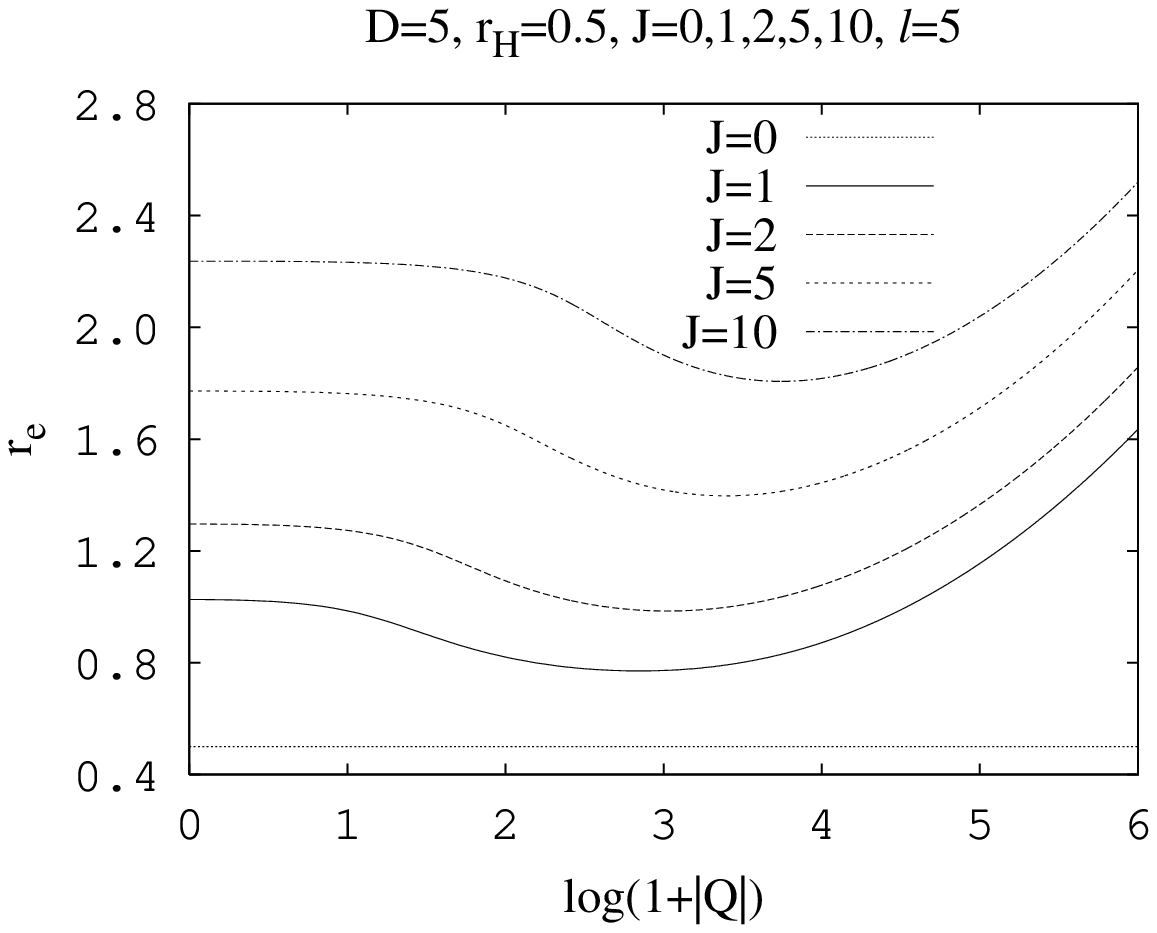}
\includegraphics[width=87mm,angle=0,keepaspectratio]{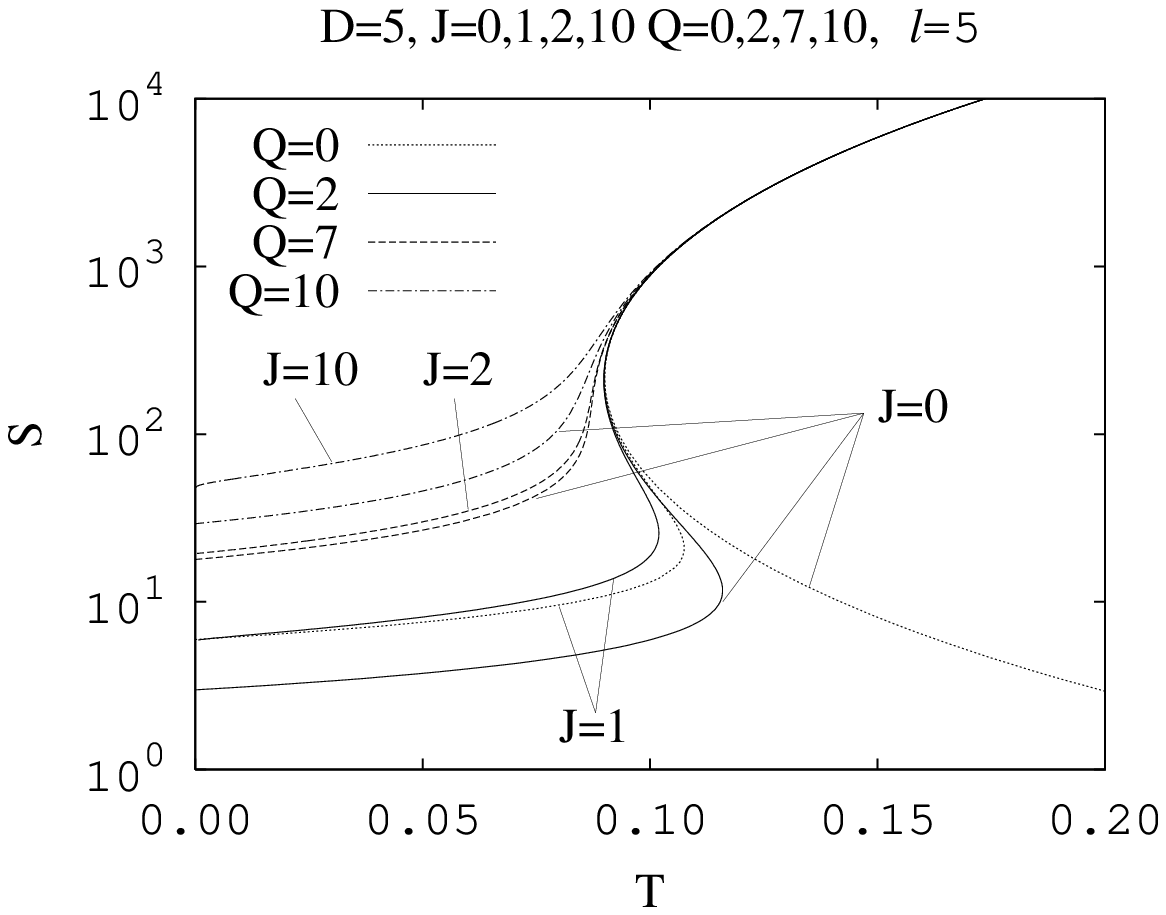}
}}}
\caption{{\small 
Left: Ergosurface (isotropic) radius $r_e$ $vs.$~the charge $Q$
for equal-magnitude angular momenta black hole solutions in $D=5$ dimensions
for fixed $r_{\rm H}=0.5$, $\ell=5$, and
several values of the angular momentum $J=0$, 1, 2, 5, 10.
Right: Entropy $S$ $vs.$~temperature $T$ 
for equal-magnitude angular momenta black hole solutions in $D=5$ dimensions
for fixed charge $Q=0, 2, 7, 10$, angular momentum $J=0, 1, 2, 10$
and $\ell=5$.}}
\end{figure}

We also exhibit the ergosurface (isotropic) radius $r_e$ 
versus the charge $Q$ in Fig.~4 (left).
The horizon area and the surface gravity of the solutions
are related to the entropy and the temperature,
respectively, $S=A_H/4G_D$ and $T=\kappa/2\pi$.
The dependence of the entropy on the temperature
is shown in Fig.~4 (right) and discussed below
with respect to the thermodynamical properties of the solutions.

\section{Further remarks}

Despite the increasing number of new interesting solutions, 
the closed form expression for the higher dimensional 
electrically charged rotating solutions in EM theory 
has not yet been obtained.
We have therefore applied a numerical approach to 
study these solutions and their properties.

Our numerical analysis indicates the existence of charged
rotating black holes in $D=2N+1$ dimensions, with $D\geq 5$,
possessing a regular horizon of spherical topology 
and $N$ equal-magnitude angular momenta. 
These solutions represent the AdS counterparts 
of asymptotically flat EM solutions,
discussed recently \cite{Kunz:2006eh,Kunz:2005nm}.

This class of solutions may provide a fertile ground for further study of
charged rotating configurations in gauged supergravity models.
For $D=5$ there should be no difficulty in principle,
using the techniques applied in \cite{Kunz:2005nm},
to extend these solutions to the general case with two
distinct angular momenta.
Higher dimensional rotating EM topological black holes 
with a horizon of negative curvature are also likely to exist
for $\Lambda<0$.

The study of the solutions discussed in this paper in an AdS/CFT context 
represents an interesting open question.
Although this may require to embed them in a supergravity model,
one should remark that for the ansatz considered here,
the boundary metric is ${\it not}$ rotating and corresponds to
a static Einstein universe in $2N$ dimensions.
The background metric upon which the dual field theory resides is
$\gamma_{ab}=\lim_{r \rightarrow \infty} \frac{\ell^2}{r^2}h_{ab}$ 
and corresponds to a $(D-1)$ 
static Einstein universe with line element
\begin{eqnarray}
\label{b-metric}
\gamma_{ab}dx^a dx^b=-dt^2+\ell^2d\Omega^2_{D-2}.
\end{eqnarray}
One can use the AdS/CFT ``dictionary'' to predict
qualitative features of a quantum field theory in this background.
For example, the expectation value of the dual CFT stress-tensor
can be calculated using the relation \cite{Myers:1999qn}
\begin{eqnarray}
\label{r1}
\sqrt{-\gamma}\gamma^{ab}<\tau_{bc}>=
\lim_{r \rightarrow \infty} \sqrt{-h} h^{ab}{\rm T}_{bc}.
\end{eqnarray}
Restricting to the five-dimensional case,
one finds (with $x^1=\theta_1,~x^2=\vphi_1,~ x^3=\vphi_2,~x^4=t$)
the following interesting form for the stress-tensor
\begin{eqnarray}
\label{st1}
8 \pi G_5 <\tau^{a}_b> =&&
\frac{1}{2\ell}\big(\frac{A}
{\ell^4}+\frac{1}{4}\big)\left( \begin{array}{cccc}
1&0&0&0
\\
0&1&0&0
\\
0&0&1&0
\\
0&0&0&-3
\end{array}
\right)
+
\frac{2B}{ \ell^5} \left( \begin{array}{cccc}
0&0&0&0
\\
0&\sin^2 \theta_1&\cos^2 \theta_1&0
\\
0&\sin^2 \theta_1&\cos^2 \theta_1&0
\\
0&0&0&-1
\end{array}
\right)
\\
\nonumber
&&+
\frac{2\hat{J}}{ \ell^3 } \left( \begin{array}{cccc}
0&0&0&0
\\
0&0&0&-\frac{1}{\ell^2}
\\
0&0&0&-\frac{1}{\ell^2}
\\
0&\sin^2 \theta_1&\cos^2 \theta_1&0
\end{array}
\right),
\end{eqnarray}
where
$A=5(\tilde{\beta}-\tilde{\alpha})$, $B=3\tilde{\alpha}-4\tilde{\beta}$.
As expected, this stress-tensor is finite and covariantly conserved. 
It is also traceless as expected from the
absence of a conformal anomaly for the static boundary metric (\ref{b-metric}) 
\cite{Skenderis:2000in}.

From the AdS/CFT correspondence, we expect these black holes to be described by
some thermal states
in a dual theory, formulated in a static Einstein universe.
Therefore it appears to be interesting to study the thermodynamics
of charged rotating EM solutions to see, how rotation affects the  
thermodynamic properties as compared to those of
the static RN-AdS solutions \cite{Chamblin:1999tk}.
This can already be attempted
based on the numerical results presented in Section 4, 
valid for solutions with Lorentzian signature.

A black hole as a thermodynamic system is unstable
if it has negative specific heat.
In the canonical ensemble, the charge and angular
momentum are fixed parameters,
and the response function whose sign determines the thermodynamic
stability is the heat capacity
$C=T\left(\frac{\partial S}{\partial T}\right)_{J,Q}.$
As is known, 
small Schwarzschild-AdS black holes ($i.e.$~$J=Q=0$) 
have negative specific heat,
but large size black holes have positive specific heat \cite{hawking1}. 
There exists a discontinuity
of the specific heat as a function of the temperature 
for some critical value of the horizon radius $r_{\rm H}$,
and thus small and large black holes are found
to be somewhat disjoint objects.

The results here indicate that this discontinuity persists 
for charged rotating black holes in EM theory, provided that
the electric charge $Q$ and the angular momentum $J$ are not very large. This
is seen in Fig.~4 (right), 
where the entropy $S$ of a set of $D=5$ EM black holes is
shown versus the temperature $T$. 
For small values of $Q$ and $J$,
the entropy increases with increasing temperature
up to a first critical point, from where it continues to increase
with now decreasing temperature up to a second critical point,
beyond which entropy and temperature both increase monotonically. 
For fixed $J$, the region where
$\frac{\partial S}{\partial T}|_{J,Q}<0$ shrinks,
as $Q$ is increased,
and there exists a critical value of $Q$ 
above which $\frac{\partial S}{\partial T}|_{J,Q}>0$. 
The analogous pattern is seen for fixed $Q$ and varying $J$. 
Thus the solutions become more thermally stable
as $Q$ and/or $J$ increase.
 

\section*{Acknowledgements}

FNL gratefully acknowledges Ministerio de Educaci\'on y Ciencia for
support under grant EX2005-0078.
The work of E.R.~was carried out in the framework of Enterprise--Ireland
Basic Science Research Project SC/2003/390.


\end{document}